\documentclass[a4paper,10pt]{article}
 \usepackage{a4}
\usepackage{epsfig}
\usepackage{amsmath}
\usepackage{amsfonts}
\usepackage{amssymb}

\newcommand{\pt}{{$p_{\rm T}$ }}
\newcommand{\vt}{{$v_{\rm 2}$ }}

\title{Photons at PHENIX}

\author{Baldo Sahlmueller
        \thanks{for the PHENIX collaboration}\\
       IKP Muenster, Germany}

\begin{document}

\maketitle


\begin{abstract}
Direct photons are a powerful probe to study the properties of the unique matter created in ultrarelativistic heavy-ion collisions. They carry information on the various stages of a heavy-ion collision. At different transverse momenta ($p_{\mathrm{T}}$), different production processes dominate the direct photon yield in heavy-ion collisions. Photons at high \pt can be used to study initial hard scattering processes while photons at low and intermediate \pt provide direct information on the hot and dense medium created in such collisions since they origin predominantly from jet-medium interactions and from thermal radiation from the medium itself. PHENIX has measured direct-photon yields over a broad \pt and energy range in different collision systems such as Au+Au and
Cu+Cu, allowing systematic studies of the behavior of direct photons in heavy-ion collisions. Two different methods have been used to measure the direct photons. An excess of direct photons in Au+Au collisions at low \pt beyond the expectation from $p$+$p$ collisions was measured for the first time via internal conversion. Furthermore, the  azimuthal anisotropy parameter \vt has been
measured. The direct photon \vt sheds light on the different processes contributing to the production of direct photons. The $p$+$p$ data at the same energy provide a baseline for understanding the heavy-ion data, but are also interesting in their own right, e.g. for testing pQCD calculations or - as RHIC collides polarized protons - for constraining models on the gluon contribution to the proton spin.
\end{abstract}


\section{Introduction}
In collisions of heavy-ions or of protons, a large number of photons can be measured. They can be roughly divided into photons from hadron decays and into direct photons which originate from several different sources \cite{Gale:2006ci}. The first source of direct photons is non-thermal and therefore also occurs in $p$+$p$ collisions. Such non-thermal photons are produced in processes described by next-to-leading order pQCD calculations, such as initial hard scatterings or bremsstrahlung emission from a scattered parton. These photons dominate the overall direct photon spectrum in heavy-ion collisions at high transverse momenta. At low transverse momenta, thermal photons \cite{Stankus:2005ab} are expected to be the dominant source. Such thermal photons are emitted by a possible quark-gluon plasma phase as well as by the hadron gas. Furthermore, photons can be emitted by the interaction of a parton jet with the medium, for example the medium can induce bremsstrahlung. Such photons are expected to play a dominant role at intermediate transverse momenta. Measuring direct photons in $p$+$p$ collisions is a good test of pQCD calculations. Furthermore, the measurement of direct photons in $p$+$p$ is an important baseline for the understanding of direct photon production in heavy-ion collisions. Through their production in quark-gluon Compton scattering, direct photons are also sensitive not only to the magnitude, but also the sign of the polarized gluon distribution
and can therefore shed light on the spin structure of protons. In heavy-ion collisions, photons are a good probe as they do not interact strongly and can thus traverse a strongly interacting medium (mostly) unaffected. As aforementioned, different production mechanisms are involved in creating the overall direct photon spectrum. The azimuthal anisotropy of direct photon production is expected to be sensitive to such different production mechanisms \cite{Gale:2006ci}, therefore it is crucial to measure both the nuclear modification factor and the azimuthal anisotropy of direct photons in heavy-ion collisions for understanding the different production mechanisms.

\section{Experimental Setup}
The PHENIX experiment \cite{Adcox:2003zm} is designed to study collisions of nuclei and of protons by looking at a large number of possible probes. The direct photon measurement presented in the following is performed within the PHENIX central spectrometer, located at both sides of the beam axis, covering $\pi$ in azimuth and $\eta=\pm 0.35$ in pseudorapidity. The detectors in the central spectrometer are arranged onion like, they can be subdivided roughly into three groups: tracking detectors, PID detectors, and calorimeters. Tracking detectors in PHENIX are the Drift Chambers (DC), the Pad Chambers (PC), and the Time Expansion Chamber (TEC). These detectors measure the tracks of charged particles and thus they play an important role in measuring electrons. Particle ID detectors are a Ring-Imaging \u{C}erenkov Detector (RICH), the Aerogel  \u{C}erenkov Counter (ACC), and the Time-of-Flight detector (TOF). The RICH separates electrons and pions below the $\pi$ \u{C}erenkov threshold of 4~GeV/$c$. The outermost detector in the PHENIX setup is an electromagnetic calorimeter (EMCal) for photon and electron measurement, consisting of eight sectors, each covering $\pi/8$ in azimuth. Six of these sectors consist of a lead scintillator sandwich calorimeters, four in the west and two in the east arm. The other two sectors in the east arm consist of a lead glass \u{C}erenkov calorimeter. Global detectors for centrality and reaction plane measurements include the Beam-Beam Counters (BBC) and the Zero-Degree Calorimeters (ZDC). A distinct reaction plane detector has been added to PHENIX in 2007.

\section{Measuring Direct Photons}
The large background of decay photons from hadrons such as $\pi^{0}$ or $\eta$ makes the measurement of direct photons difficult. Two different methods have been used at PHENIX to extract the direct photon signal from the decay photon background: the so-called subtraction method, and the so-called internal conversion method. The subtraction or statistical method has its advantages at high transverse momenta but fails to extract a significant direct photon yield at low \pt due to large systematic uncertainties. It has been well established in different analyses \cite{Adler:2005ig, Adler:2006yt}. The idea of this method is to subtract the decay photons from the different hadronic decays from all measured photons. Photon candidates are measured with the electromagnetic calorimeter, the hadron and lepton contribution to this spectrum is subtracted and the spectrum is corrected for detector effects such as the geometrical acceptance and the reconstruction efficiency, leading to a spectrum of inclusive photons $\gamma_{\mathrm{inclusive}}$. The spectrum of decay photons is obtained via simulations, taking the measured $\pi^{0}$ spectrum \cite{Adare:2007dg, Adare:2008cx, Adare:2008qa} as input, and using $m_{T}$ scaling for the other mesons such as $\eta$ or $\omega$. The decay photons are then not subtracted directly but via the so-called double ratio $R_{\gamma} = \frac{\gamma/\pi^{0}_{\mathrm{data}}}{\gamma/\pi^{0}_{\mathrm{decay}}}$. In this ratio, common systematic uncertainties on the inclusive photon and the $\pi^{0}$ spectra cancel, the direct photon yield can eventually be calculated as $\gamma_{\mathrm{dir}} = \left(1-\frac{1}{R_{\gamma}}\right)\cdot \gamma_{\mathrm{inclusive}}$. This method has also been applied in Au+Au collisions to reaction-plane dependent data to obtain the azimuthal anisotropy parameter \vt for direct photons, and in collisions of polarized proton beams it was used for the measurement of the double longitudinal spin asymmetry $A_{\mathrm{LL}}$. At low transverse momenta, the method cannot be applied as the remaining systematic uncertainties are larger than the small signal. Furthermore, at high transverse momenta, the merging of $\pi^{0}$ decay photons has to be taken into account.\\

\begin{figure}[htp]
 \centering
 \includegraphics[width=12cm]{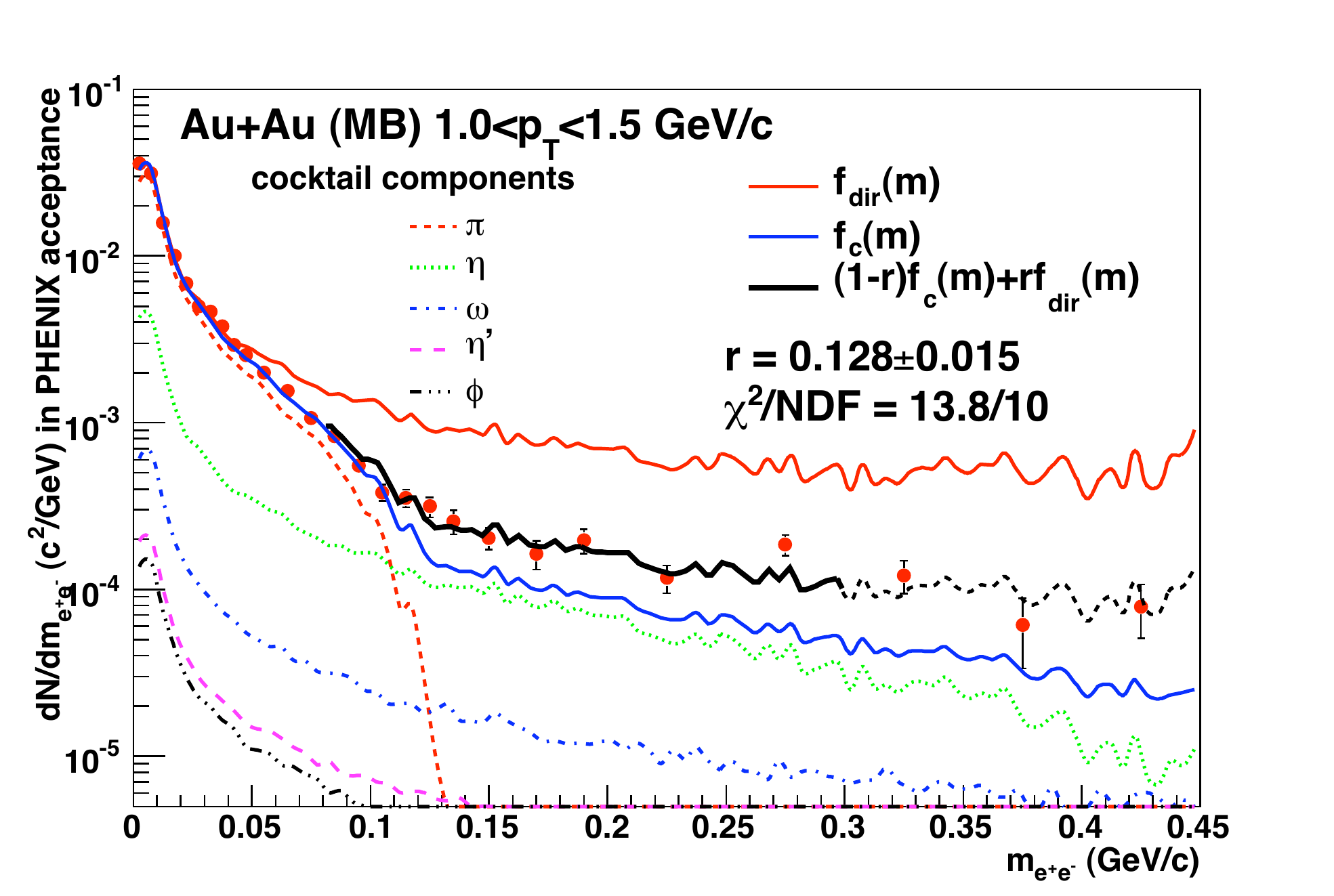}
 \caption{Mass distribution of $e^{+}e^{-}$ pairs for Au+Au minimum bias events for 1$<$\pt$<$1.5~GeV/$c$. The range of the two-component fit (as explained in the text) is $80<m_{ee}<300$ MeV/$c^{2}$. The black dashed curve shows $f(m_{ee})$ outside the fit range. The mentioned excess over the hadron cocktail can also be seen, comparing the blue $f_{c}(m_{ee})$ with the data points.}
 \label{fig:massdistro}
\end{figure}

The internal conversion method \cite{Adare:2008fqa} is based on the assumption that every source emitting real photons also emits virtual photons that subsequently convert into low mass electron-positron pairs. Such internal conversion occurs for example in the $\pi^{0}$ Dalitz decay, or when a source emits virtual photons directly. The mass distribution of such photons can be described by
\begin{eqnarray}
\frac{d^{2}n_{ee}}{dm_{ee}}=\frac{2\alpha}{3\pi}\frac{1}{m_{ee}}\sqrt{1-\frac{4m_{e}^{2}}{m_{ee}^{2}}}\cdot(1+\frac{2m_{e}^{2}}{m_{ee}^{2}})Sdn_{\gamma}\mbox{    .}
\end{eqnarray}
$\alpha$ is the fine structure constant, $S$ is a process dependent form factor, $S=1$ for direct photons in case $m\rightarrow 0$, or $m \ll p_{\mathrm{T}}$, while $S = |F(m_{ee}^{2})|^{2}(1-\frac{m_{ee}^{2}}{M_{h}^{2}})^{3}$ for the Dalitz decay of $\pi^{0}$ an $\eta$. Therefore, the factor $S = 0$ for $m_{ee}>M_{h}$, this cut-off can be used to separate the direct photon signal from the background.\\
In a first analysis step, all $e^{+}e^{-}$ pairs are measured in a certain \pt region, a cut removes pairs from conversion in the detector material, while the uncorrelated background is calculated via event mixing and the correlated background is removed based on like-sign pairs. The mass distribution is then corrected for the electron reconstruction efficiency and - in the $p$+$p$ case - for the trigger efficiency. The resulting mass distribution is compared to the expectation from a simulated cocktail of hadron decays. An excess over this expectation can be seen in Au+Au collisions as well as - though smaller - in $p$+$p$ collisions. A possible source for such an excess are internal conversions of direct photons. In a next step, the mass distribution is fit with a two component function $f(m_{ee}) = (1-r)f_{c}(m_{ee}) + rf_{dir}(m_{ee})$, with $f_{c}(m_{ee})$ being the shape of the cocktail mass distribution and $f_{dir}(m_{ee})$ as the expected shape of the direct photon internal conversion. Both single functions are normalized to the data in the low mass region $m_{ee} < 30$ MeV/$c^{2}$ where their shape is nearly identical. An example of the mass distribution together with fits for the different contributions is depicted in Fig. \ref{fig:massdistro}. The fit is done for each \pt bin within several mass ranges $m_{\mathrm{low}} < m_{ee} < 300$~MeV/$c^{2}$, $r$ is the only fit parameter and is the ratio of direct photons and inclusive photons.

\section{Results}

\begin{figure}[htp]
 \centering
 \includegraphics[width=11cm]{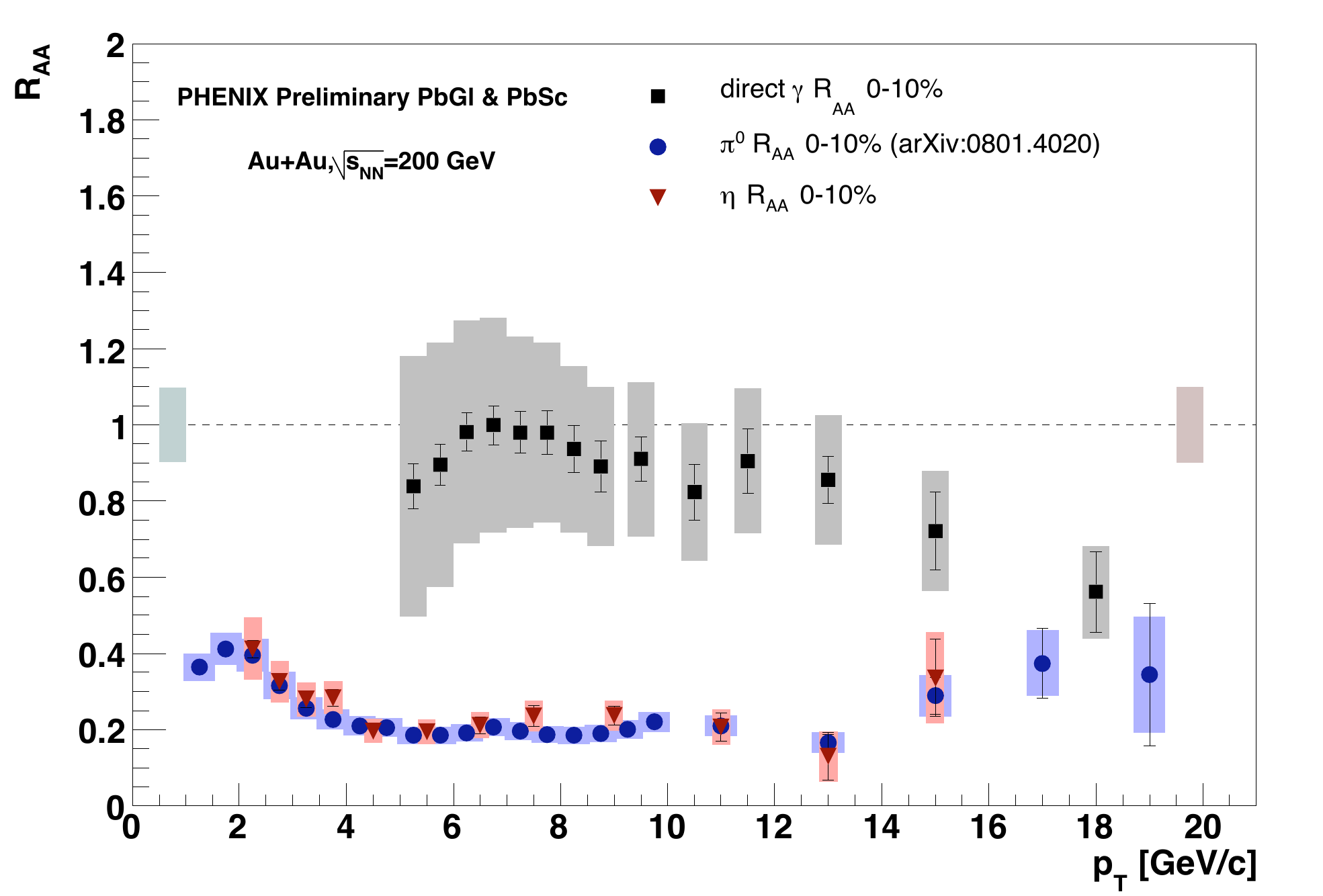}
 \caption{The nuclear modification factor $R_{\mathrm{AA}}$ for direct photons, $\pi^{0}$, and $\eta$ in Au+Au collisions at $\sqrt{s_{\mathrm{NN}}}=200$~GeV. The error bars depict \pt uncorrelated errors while the error boxes show \pt correlated systematic errors.}
 \label{fig:raaauau}
\end{figure}

The spectra measured in $p$+$p$ and Au+Au collisions have been extended to $p_{\mathrm{T}}=22$~GeV/$c$ and $p_{\mathrm{T}}=18$~GeV/$c$, respectively. These spectra can be used to calculate the so-called nuclear modification factor $R_{\mathrm{AA}}$, a measure to quantify possible nuclear effects. The nuclear modification factor of direct photons in Au+Au collisions at $\sqrt{s_{\mathrm{NN}}}=200$~GeV is shown in Fig. \ref{fig:raaauau} together with the $R_{\mathrm{AA}}$ of $\pi^{0}$ and $\eta$. In contrast to the mesons, direct photons are not suppressed at $p_{\mathrm{T}} \lesssim 15$~GeV/$c$, but show an apparent suppression at highest transverse momenta. This suppression of direct photons is not yet fully understood. One possible explanation could be the isospin effect, a difference in the different cross sections $\sigma^{in}_{pp}$ and $\sigma^{in}_{pn}$, leading to different cross sections for colliding protons or neutrons. To further study this isospin effect, one can look into other collision systems such as Cu+Cu. 


\begin{figure}[htp]
 \centering
 \includegraphics[width=8.5cm]{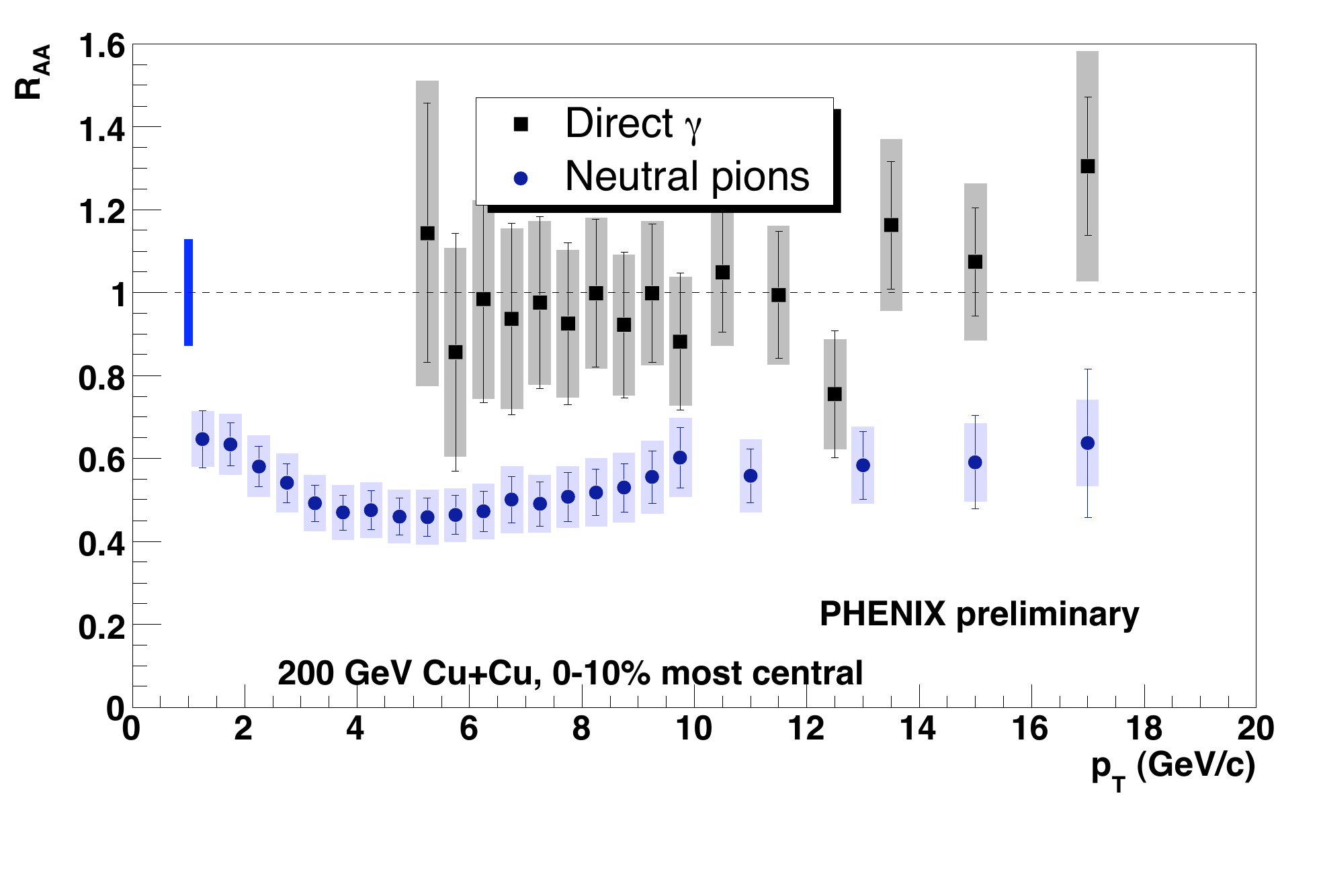}
 \caption{The nuclear modification factor $R_{\mathrm{AA}}$ for direct photons and $\pi^{0}$ in Cu+Cu collisions at $\sqrt{s_{\mathrm{NN}}}=200$~GeV. The error bars depict \pt uncorrelated errors while the error boxes show \pt correlated systematic errors.}
 \label{fig:raacucu}
  \end{figure}
  
 \begin{figure}[hpt]
   \centering
  \includegraphics[width=8.5cm]{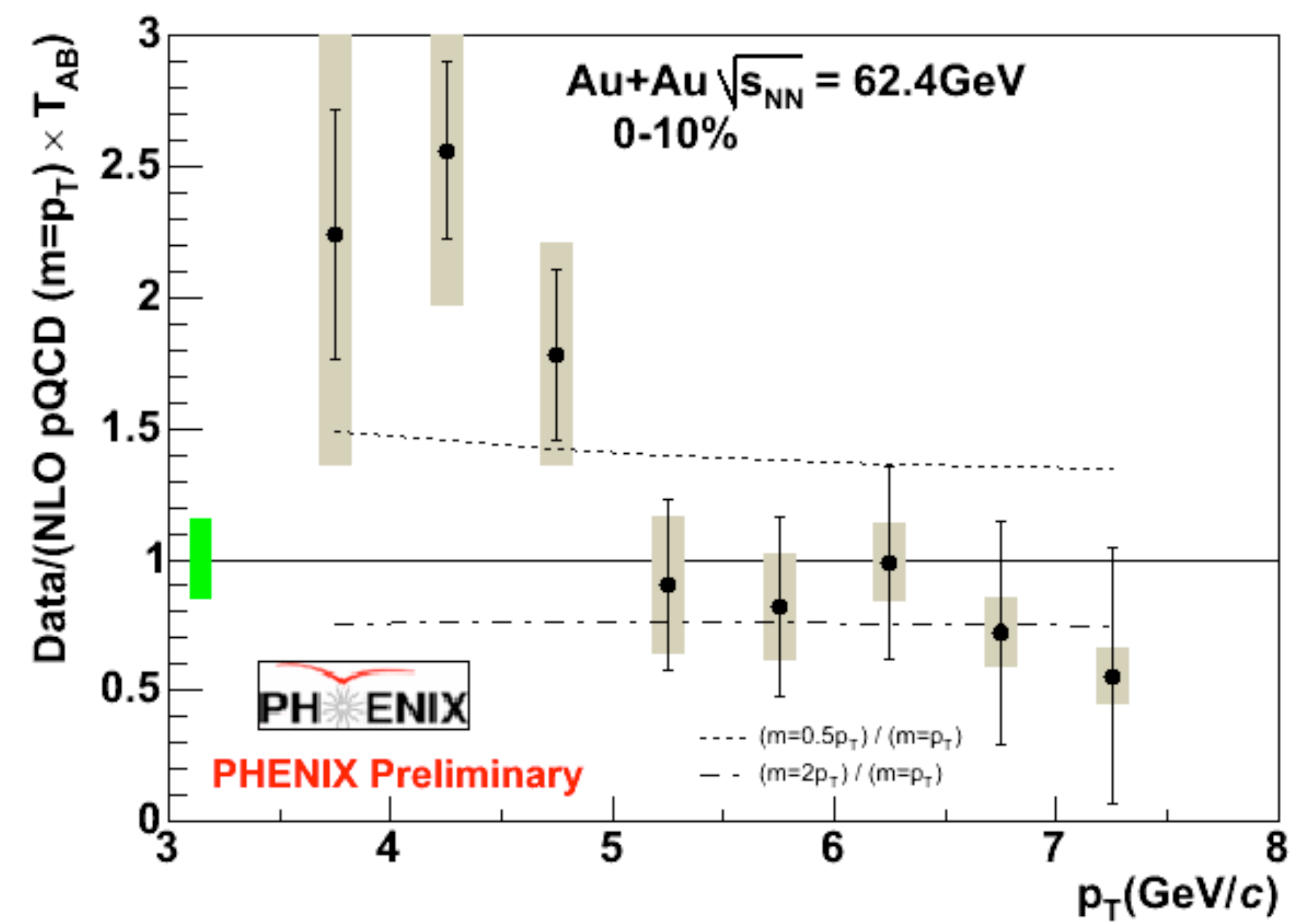}
 \caption{The direct photon $R_{\mathrm{AA}}$ as a function of \pt  in Au+Au collisions at $\sqrt{s_{\mathrm{NN}}}=62.4$~GeV. The error bars depict \pt uncorrelated errors while the error boxes show \pt correlated systematic errors.}
 \label{fig:raaauau62}
\end{figure}

PHENIX has measured the direct photon $R_{\mathrm{AA}}$ in such collisions at $\sqrt{s_{\mathrm{NN}}}=200$~GeV for  $p_{\mathrm{T}} \leq 17$~GeV/$c$, the result is shown in Fig. \ref{fig:raacucu} together with the $\pi^{0}$ $R_{\mathrm{AA}}$. No suppression has been observed for direct photons in such collisions, but the \pt range is less than in Au+Au and the uncertainties are larger at highest \pt. As the isospin effect scales with $x_{\mathrm{T}} = 2p_{\mathrm{T}}/\sqrt{s}$, looking at lower collision energies would make studies on this effect available at lower transverse momenta where the merging of $\pi^{0}$ decay photons on the calorimeter does not yet play a role. Therefore, direct photons in Au+Au collisions at $\sqrt{s_{\mathrm{NN}}}=62.4$~GeV have been measured as well, the nuclear modification factor is shown in Fig. \ref{fig:raaauau62}. Unfortunately, the uncertainties of the measurement are too large to draw any conclusions.\\

\begin{figure}[htp]
\centering
 \includegraphics[width=7.8cm]{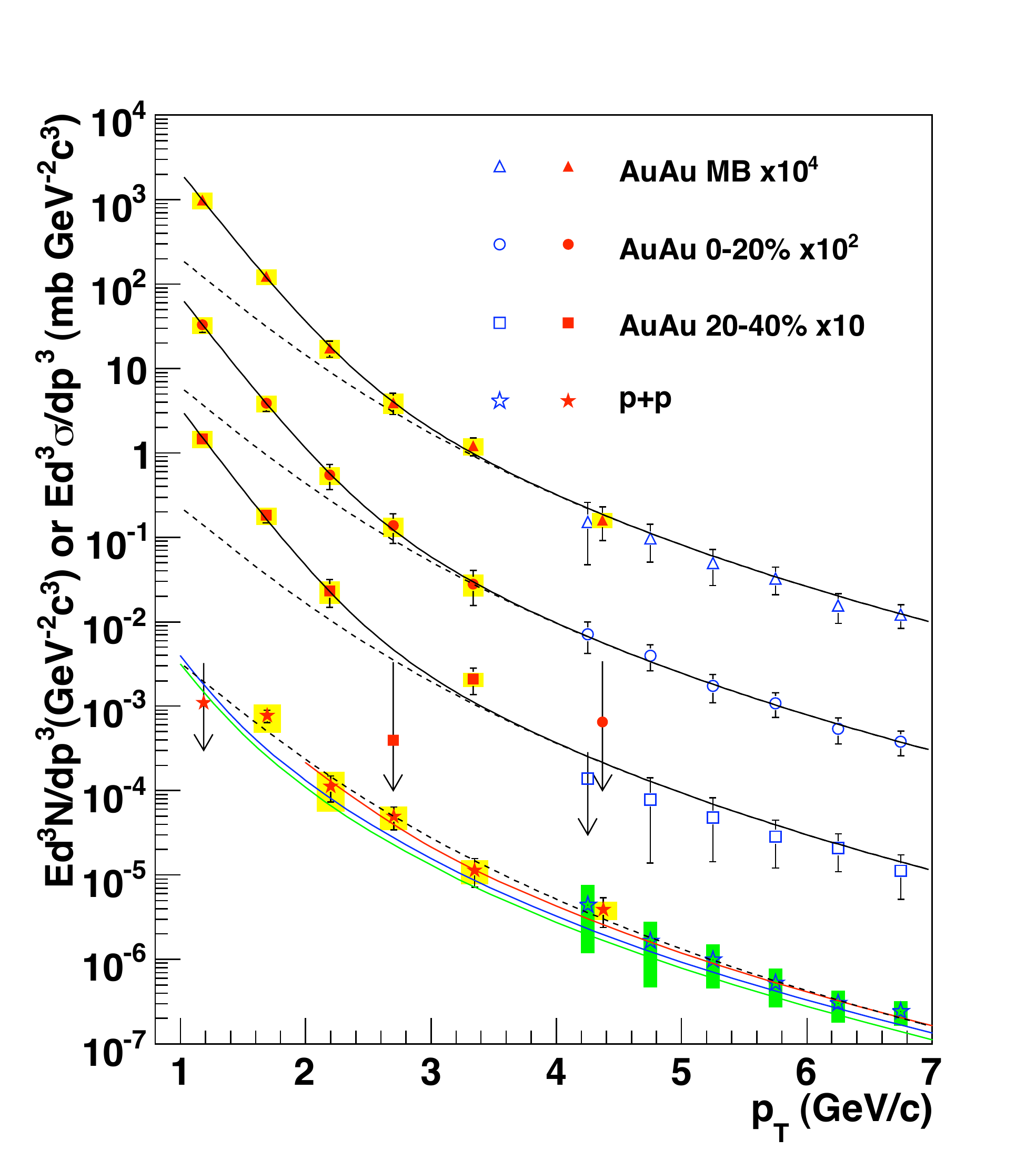}
 \caption{The direct photon invariant cross section ($p$+$p$) and invariant yield (Au+Au) as a function of $p_{\mathrm{T}}$. The filled points are from \cite{Adare:2008fqa}, the open points from \cite{Adler:2005ig}, \cite{Adler:2006yt}. The three curves on the $p$+$p$ data are from NLO pQCD calculations \cite{Gordon:1993ab}, the dashed lines are a modified power-law fit to $p$+$p$, scaled by $T_{\mathrm{AA}}$. The black curves on the Au+Au data are the $T_{\mathrm{AA}}$ scaled $p$+$p$ fit plus an exponential.}
 \label{fig:dirgam_low}
\end{figure}

At low transverse momenta (\pt $< 4.5$~GeV/$c$), PHENIX has measured the ratio $r=\gamma_{\mathrm{direct}}/\gamma_{\mathrm{inclusive}}$ in $p$+$p$ and Au+Au collisions at  $\sqrt{s_{\mathrm{NN}}}=200$~GeV via internal conversion. This ratio is consistent with pQCD predictions \cite{Gordon:1993ab} in $p$+$p$ collisions while the experimental result exceeds a binary-scaled pQCD prediction in Au+Au collisions significantly. The direct photon invariant yield can then be calculated as $dN^{\mathrm{direct}}(p_{\mathrm{T}}) = r \times dN^{\mathrm{inclusive}}(p_{\mathrm{T}})$. Fig. \ref{fig:dirgam_low} shows the direct photon invariant cross section and the invariant yield for $p$+$p$ and Au+Au collisions, respectively. For Au+Au collisions, different centrality selections are shown. The results obtained with the statistical method are added to the plot as well, showing a good agreement with the internal conversion data in the overlap region. The pQCD calculation is consistent with the $p$+$p$ data within uncertainties for $p_{\mathrm{T}} > 2$~GeV/$c$. The same data can also be well described by a modified power law function $(A_{pp}(1+2p_{\mathrm{T}}^{2}/b)^{-n})$ which is represented by the dashed line in the figure. The $p$+$p$ curve, scaled by $T_{\mathrm{AA}}$, the nuclear overlap function, is significantly below the Au+Au data for $p_{\mathrm{T}} < 2.5$~GeV/$c$. The Au+Au data have thus been fit with another function, an exponential plus the $T_{\mathrm{AA}}$ scaled $p$+$p$ fit function $(Ae^{-p_{\mathrm{T}}/T}+T_{\mathrm{AA}}\times(A_{pp}(1+2p_{\mathrm{T}}^{2}/b)^{-n})$. The only free parameters in the fit are $A$ and $T$, the inverse slope of the exponential. For the 20\% most central events, $T$ has been found to be $T = 221 \pm 23 \pm 18$~MeV. If the direct photon excess is of thermal origin, $T$ is related to the initial temperature $T_{\mathrm{init}}$ of the medium, in hydrodynamical models~\cite{d'Enterria:2005vz}, $T_{\mathrm{init}}$ is 1.5 to 3 times $T$. The various models differ in $T_{\mathrm{init}}$ and thermalization time $\tau_{0}$, these two parameters appear to be related, such that $T_{\mathrm{init}}\cdot\tau_{0}\approx\mathrm{const}$.\\

\begin{figure}[htp]
 \centering
 \includegraphics[width=10cm]{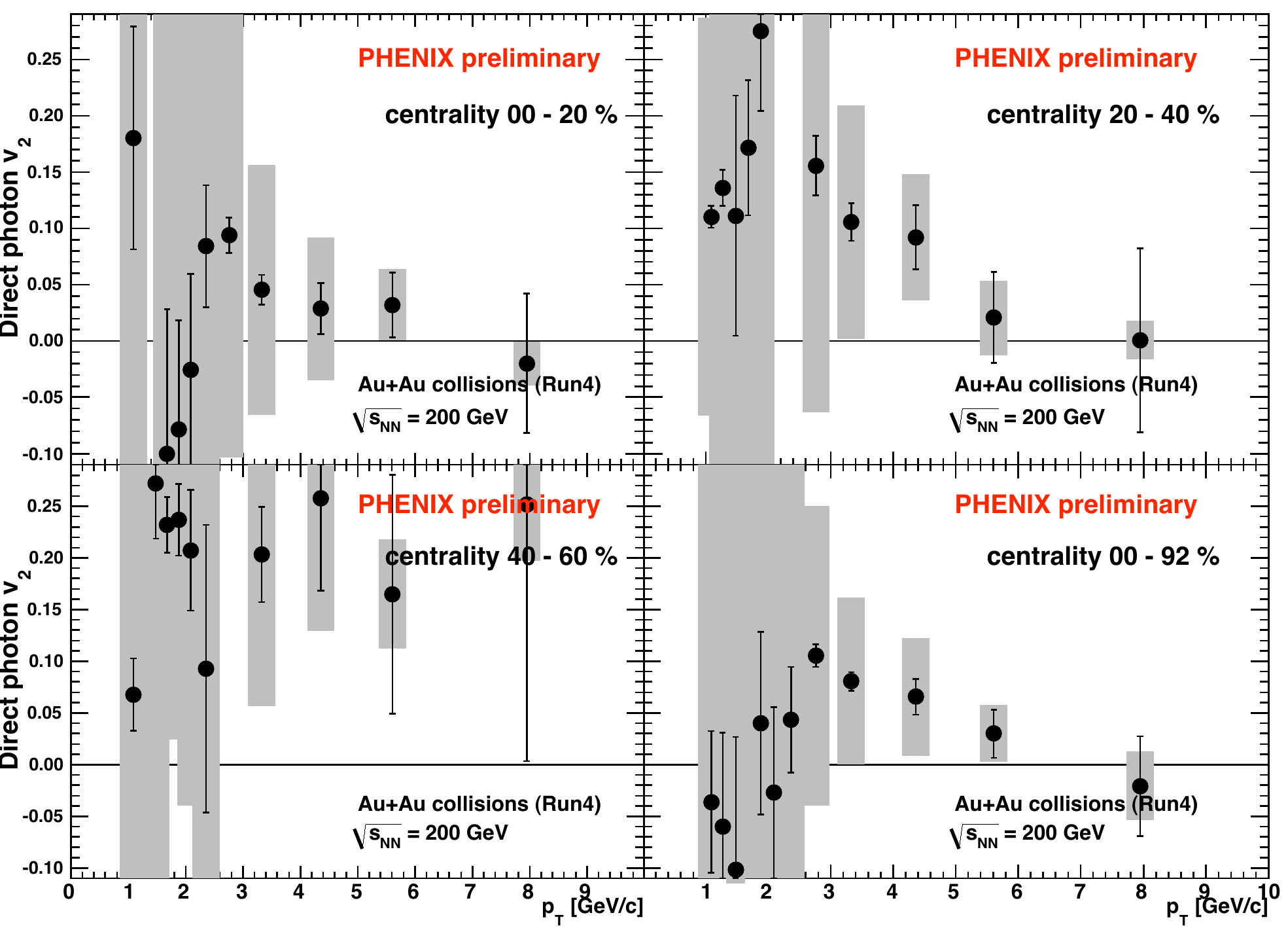}
 \caption{The direct photon azimuthal anisotropy parameter \vt as a function of \pt for different centrality selections. The error bars depict \pt uncorrelated errors while the error boxes show \pt correlated systematic errors.}
 \label{fig:v2}
\end{figure}

The direct photon \vt has been measured with the same Au+Au data set as the aforementioned direct photon $R_{\mathrm{AA}}$ in Fig. \ref{fig:raaauau}.  The result is shown in Fig. \ref{fig:v2} for different centrality selections. Within the uncertainties of the measurement, the data are consistent with a zero \vt though the data points might slightly favor a positive \vt. But so far, the uncertainties are too large to favor or disregard theoretical models.\\

\begin{figure}[htp]
 \centering
 \includegraphics[width=8cm]{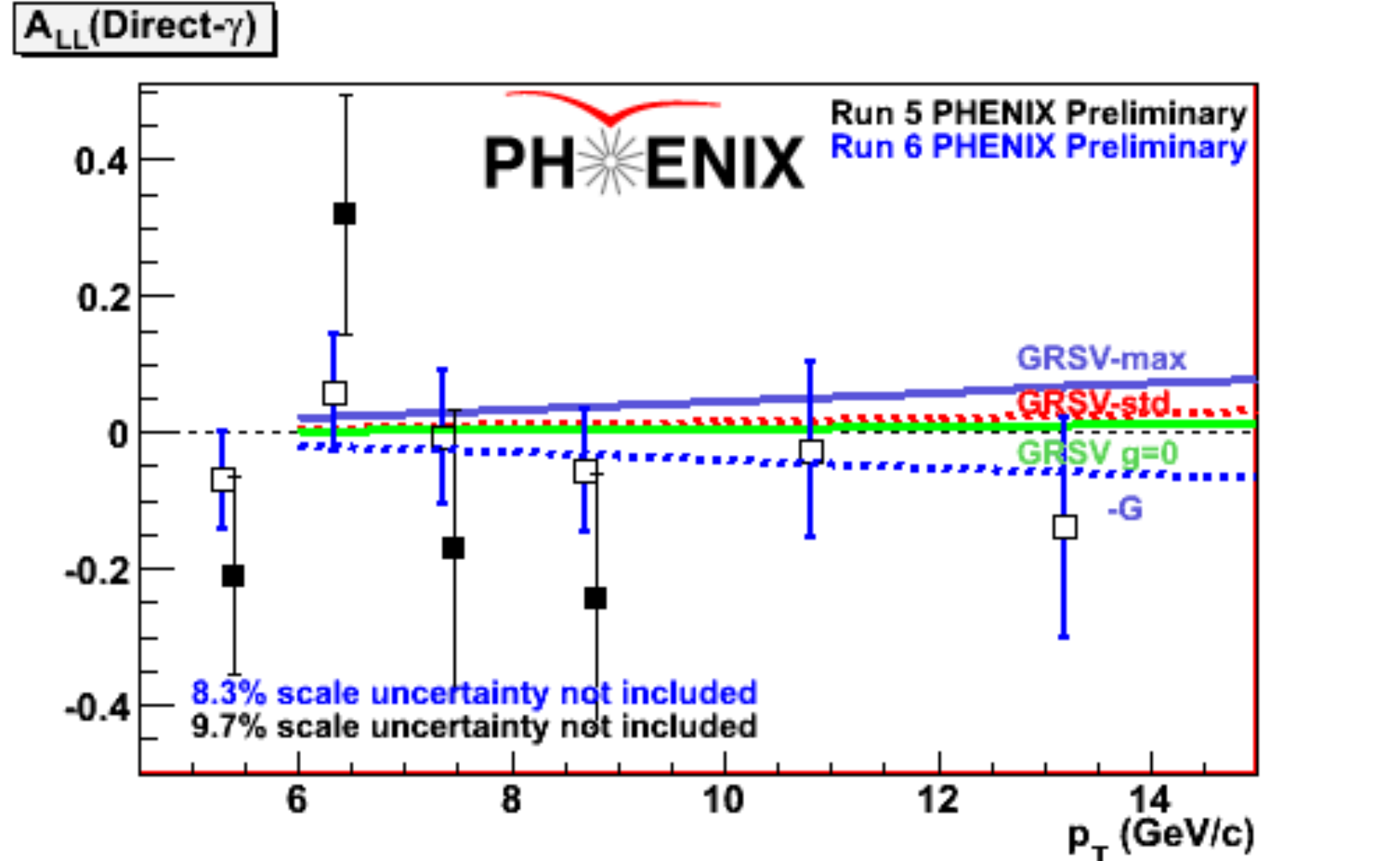}
 \caption{The direct photon longitudinal spin asymmetry $A_{\mathrm{LL}}$ in $p$+$p$ collisions at  $\sqrt{s_{\mathrm{NN}}}=200$~GeV. The data from two different RHIC runs are shown together with theoretical expectations for different gluon polarizations $\Delta$G \cite{Gluck:2000dy}.}
 \label{fig:all}
\end{figure}

In $p$+$p$ collisions, the double longitudinal spin asymmetry $A_{\mathrm{LL}}$ has been measured in data sets from two different PHENIX runs. The results are shown in Fig. \ref{fig:all} together with theoretical calculations for different gluon polarizations $\Delta$G \cite{Gluck:2000dy}. The 2006 measurement significantly improves the accuracy of the result, but still the uncertainties of the measurement are too large to constrain~$\Delta$G.

\section{Conclusion}

Direct photons have been measured by the PHENIX experiment in different collision systems at different energies. Two different methods have been used to extract the direct photon signal from the large background of decay photons. Both have their advantages in a certain transverse momentum range. The statistical method is more feasible at high \pt, while the internal conversion method allows a direct photon measurement at low \pt. The statistical method has been extended to measure asymmetries in the direct photon production, the parameter \vt has been measured in Au+Au collisions, and the longitudinal spin asymmetry $A_{\mathrm{LL}}$ has been measured in $p$+$p$ collisions.\\
The nuclear modification factor $R_{\mathrm{AA}}$ is consistent with unity for central Cu+Cu and Au+Au collisions at 200~GeV/$c$. In Au+Au collisions, an apparent suppression of the direct photon signal sets in at $p_{\mathrm{T}}\gtrsim$~15~GeV/$c$, an observation not yet understood. The measurement of direct photons at low \pt shows an excess in Au+Au collisions above a scaled $p$+$p$ measurement. An exponential fit to this excess has an inverse slope of $T = 221 \pm 23 \pm 18$~MeV which is consistent with hydrodynamic models with $T_{\mathrm{init}} \approx 300 - 600$~MeV.\\
The measurements of both \vt and $A_{\mathrm{LL}}$ so far suffer from large uncertainties and cannot yet be used to support or disregard theoretical models. However, with additional detectors and more data taken in later RHIC runs, it is likely to reduce the size of the uncertainties and get more significant results for those measurements.

\end{document}